\documentclass[aps,prd,
nofootinbib,
preprintnumbers,
twocolumn,
]{revtex4-1}

\usepackage{amsmath}
\usepackage{amssymb}
\usepackage{graphicx}
\usepackage{bm}
\usepackage[colorlinks,linktocpage]{hyperref}
\hypersetup{colorlinks=true, citecolor=blue}

\usepackage[normalem]{ulem} 

\usepackage{color}

\begin{document}
\preprint{INR-TH-2021-028}

\title{Solar mass black holes from neutron stars and bosonic dark matter}
\date{25/02/2022}
\author{Raghuveer Garani}
\affiliation{INFN Sezione di Firenze, Via G. Sansone 1, I-50019 Sesto
  Fiorentino, Italy}
\author{Dmitry Levkov}
\affiliation{Institute for Nuclear Research of the Russian Academy
  of Sciences, Moscow 117312, Russia}
\affiliation{Institute for Theoretical and Mathematical Physics, MSU,
  Moscow 119991, Russia}
\author{Peter Tinyakov}
\affiliation{Service de Physique Th\'{e}orique, Universit\'{e} Libre
  de Bruxelles (ULB),\\CP225 Boulevard du Triomphe, B-1050 Bruxelles,
  Belgium}

\begin{abstract} 
  Black holes with masses $\approx 1\, M_{\odot}$ cannot be produced via
  stellar evolution. A popular scenario of their formation
  involves transmutation of neutron stars~--- by accumulation of dark
  matter triggering gravitational collapse in the star centers. We
  show that this scenario can be realized in the models of bosonic
  dark matter despite the apparently contradicting requirements on the interactions of
  dark matter particles: on the one hand, they 
  should couple to neutrons strongly enough to be captured inside
  the neutron stars, on the other, their loop--induced
  self--interactions impede collapse. Observing that these
  conflicting conditions are imposed at different scales, we
  demonstrate that models
  with efficient accumulation of dark matter can be deformed
  at large fields to make unavoidable its subsequent 
  collapse into a black hole. Workable examples
  include weakly coupled models with bended infinite valleys.
\end{abstract}

\maketitle

\section{Introduction}
\label{sec:introduction}
The existing gravitational wave detectors are sensitive to compact
objects~---  black holes (BHs) and neutron stars (NSs)~--- of masses
ranging from tens of solar masses  down to $M_\odot$, 
see~\cite{LIGOScientific:2017vwq, LIGOScientific:2021job}. Thus far,
few tens of coalescing BHs and several BH--NS systems were 
detected~\cite{LIGOScientific:2018mvr, LIGOScientific:2020ibl,
  LIGOScientific:2021qlt, LIGOScientific:2021djp}, while hundreds of
observations are further expected in the near future. This will be
sufficient to map out the mass function of the stellar--size BHs.

Standard stellar evolution predicts no BHs lighter than the
heaviest neutron stars, as the Fermi pressure stabilizes
the neutron cores at masses below~${2.5 M_\odot}$,
cf.~\cite{Rhoades:1974fn, Lattimer:2012nd, Ozel:2016oaf}. The BH mass
function is hence expected to have a gap below this value, and any
detection in that region would automatically imply existence of an
exotic mechanism for black hole formation. We will refer to such light
BHs as ``solar mass black holes.''

A widely discussed possibility is production of solar mass BHs in  
the early Universe  from small--scale density perturbations. It has a
natural extension: these objects may also play 
the role of dark matter (DM). The latter hypothesis, however, is
strongly constrained by lensing~\cite{EROS-2:2006ryy}, CMB
measurements~\cite{Ricotti:2007au, Ali-Haimoud:2016mbv}, and even the 
gravitational wave observations themselves~\cite{Sasaki:2018dmp}.

Alternatively, the solar mass BHs may 
appear in the present
Universe due to neutron star collapses caused  by seed
BHs~\cite{Goldman:1989nd}. The latter BHs may in turn be of
primordial origin, and be small enough to evade the lensing and CMB
constraints. This option, however, still relies on cosmology to produce
a sufficient seed BH population.

A more intriguing possibility is the formation of
 seed BHs through the gravitational collapse of dark matter accumulated by old
neutron stars~\cite{Kouvaris:2010jy, Kouvaris:2011fi,
  McDermott:2011jp, Bell:2013xk, Kouvaris:2018wnh, Garani:2018kkd,
  Dasgupta:2020mqg}. A sufficient accumulation is possible only
if the DM is asymmetric~\cite{Davoudiasl:2012uw} and does not annihilate
in the stellar core. In the simplest case one assumes that the
DM particles carry a conserved global charge ensuring their stability,
and their antiparticles disappear during the early stages of cosmological evolution.

In the present paper we focus on the last scenario. Our goal is to
identify DM models where such catalyzed transmutation of
the neutron stars into the solar mass BHs can take place. The
mechanism we consider involves four stages: (i)~capture of DM by the
neutron star; (ii)~its thermalization; (iii)~concentration in the star
center and  gravitational collapse into a seed BH; (iv)~accretion of
the star material onto the BH and formation of the solar mass
BH. Given the  DM abundance, the stage (i) 
sets the maximum mass of DM that can be accumulated by the neutron
star. On the other hand, the stage (iii) requires some minimum amount
of DM for the successful gravitational collapse. The viability of the
overall scenario depends on the possibility to  satisfy both
requirements in the same DM 
model.

Stages (i), (ii), and (iv), i.e.\ the beginning of the process and its
very end, have been extensively discussed in the literature, see the
review~\cite{Tinyakov:2021lnt} and~\cite{Kouvaris:2013kra,
  Genolini:2020ejw}. We will provide a short summary of these results
in the next section. The main part of this paper considers 
model--dependent evolution and gravitational collapse of 
dense dark matter cloud at stage~(iii).

The minimum number of particles required for collapse is different in
the cases of bosonic and fermionic dark matter, and  crucially
depends on the DM self--interactions. In the ensemble of free DM
fermions, the Fermi pressure balances self--gravity and halts the
gravitational collapse unless the total particle number exceeds the
(analog of the) Chandrasekhar limit~\cite{Chandrasekhar:1931ih}, 
\begin{equation}
\text{free fermions:}\quad N_{cr} \sim \left({M_{\rm Pl} \over m }\right)^3 
\sim 10^{57} \left({{\rm GeV} \over m }\right)^3\,,
\label{eq:fermions}
\end{equation}
where $m$ is the DM mass. This number is prohibitively
large unless $m$ is in the 100~TeV range or
above~\cite{Kouvaris:2018wnh}. While not 
necessarily impossible, such scenarios with super-heavy asymmetric dark
matter are outside of the mainstream cosmological models, and we do
not consider them here.

If the DM is bosonic and non--interacting, the gravitational attraction of
its particles is balanced by the kinetic (``quantum'') pressure
guaranteed by the
uncertainty principle. Gravitational collapse happens if the particle
number exceeds the critical value
\begin{equation}
\text{free bosons:}\quad N_{cr}\sim \left({M_{\rm Pl} \over m }\right)^2 
\sim 10^{38} \left({{\rm GeV} \over m }\right)^2\,,
\label{eq:bosons}
\end{equation}
which is parametrically smaller than in the case of fermions. 
Bosonic DM is therefore a more promising candidate for the creation of
seed BHs. In the rest of this paper we will focus on bosons.  

While looking advantageous in the free case,
Eq.~(\ref{eq:bosons}), however, can be ruined  by
very weak DM self--interactions which are generically present in
our scenario. Indeed, DM capture and thermalization at stages  (i) and  
(ii) are possible only for sufficiently strong DM--nucleon
interactions which in turn induce DM self--couplings via
loops~\cite{Bell:2013xk}. In the simple one--field models, the induced
terms in the DM potential are quartic in fields and should be 
positive, i.e.\ repulsive, for vacuum stability. In effect, they
generically oppose the gravitational collapse and 
increase the required critical DM multiplicity to unacceptably large values. We
have therefore conflicting conditions on the DM model.

In this paper we show that the conflict can be resolved. Namely, the
number of DM particles required for collapse  can be almost as
small as in Eq.~(\ref{eq:bosons}), and all of these particles 
can be accumulated inside the neutron star by scattering off
nucleons. We start by clarifying and quantifying the requirements on
the DM models. Then we propose a generic mechanism to satisfy
them simultaneously. We find that the
smallest DM multiplicity for collapse  is achieved in the models where:
(i)~the DM potential includes a long 
valley extending to large, in some cases Planckian fields, and the
  cutoff of this potential is high enough; (ii)~both attractive
and repulsive self--interactions are suppressed along this valley;
(iii)~loop quantum corrections from interactions with the visible
sector do not break the condition~(ii).

Our mechanism is based on the fact that the conflicting requirements
are imposed at different scales. The DM capture depends on the physics in the
vicinity of  vacuum, while self--interactions obstructing the gravitational
collapse should vanish at strong fields. Thus, one can make the
potential valley {\it bended} in the field space. It may start
going along
the dark matter field $\phi$, then take a turn at some  $|\phi|
\gtrsim \Lambda$ and continue in the direction of another
field~$\chi$. In this case the interaction of $\phi$--particles with
nucleons, as we will argue, does not generate an effective  potential
for~$\chi$, and the latter may grow and collapse  into a BH
  almost as in the free bosonic case.
  
The rest of this paper is organized as follows. We review accumulation
and thermalization of DM inside the neutron stars in
Sec.~\ref{sec:generalities}. General requirements on the DM models
with the smallest critical multiplicities for gravitational collapse
are derived in Sec.~\ref{sec:grow-bose-einst}. In 
Sec~\ref{sec:models-with-bended} we propose a mechanism to satisfy
these requirements. We conclude in Sec.~\ref{sec:conclusions}. 


\section{DM inside the neutron star}
\label{sec:generalities}

In this section we summarize the existing results on DM capture and
thermalization in the neutron star. Two points will be essential for us:
(i)~the total amount of captured DM and its dependence on the strength
of DM--neutron interactions; (ii)~the fact that with time this DM forms a
Bose--Einstein condensate described by classical fields. 

For concreteness we will assume that the DM particles are globally
charged scalars with mass $m$ belonging to the
typical WIMP range from GeV to a few~TeV.

\subsection{DM capture} 
\label{sec:dm-capture}

Two generic mechanisms trap DM inside the neutron star:
accumulation during the star lifetime and gravitational capture at the
star formation. These mechanisms are cumulative. Potentially, they may
provide comparable amounts of trapped DM.

In the next sections we  consider accumulation of DM particles from the
ambient galactic halo~\cite{Press:1985ug, Gould:1987ww}. Far away from the
neutron star, the distribution of DM velocities is nearly Maxwellian with
small dispersion ${\bar v}^2$, e.g.\ $\bar{v} \sim 10^{-3}$ in the Milky
Way. But the particles acquire semi--relativistic speeds~${\sim 0.5}$
as they fall
into the neutron star. With masses in the WIMP range, they lose
energies of order~$m_n$~--- the neutron's rest mass~--- in collisions with
neutrons. This is much larger than the asymptotic energies of the particles;
thus, most of them bind gravitationally to the neutron star after the first
collision. Besides, the momentum transfer in their collisions
marginally
exceeds the Fermi momentum of neutrons, so that the degeneracy 
of the latter does not play a crucial role. Neglecting the degeneracy
is a crude approximation that overestimates the capture
rate~\cite{Bell:2020jou, Bell:2020lmm, Anzuini:2021lnv}, but in view
of comparable astrophysical uncertainties we will use it for
simplicity. Finally, we will ignore general relativity effects which
enhance the capture rate by an  order~1 number, see Ref.~\cite{Kouvaris:2007ay}.

Under these assumptions, the capture rate, i.e.\ the number of
trapped DM particles per unit time, takes the form~\cite{Press:1985ug,
Gould:1987ww},
\begin{equation}
  \frac{dN}{dt}\approx \sqrt{24 \pi} \,G \;  {\rho_{\rm DM}\over m \bar v} \,
  M_* R_* f\,.
\label{eq:capture_gen}
\end{equation}
Here $\rho_{\rm DM}$ is  ambient DM density, $M_*$ and $R_*$
represent the mass and the radius of the neutron star, and we introduced the
probability $f\leq 1$  for the DM particle to scatter during one pass
through the star. The value of $f$ is proportional to the DM--neutron 
cross section~$\sigma$,
\begin{equation}
f= {\sigma/\sigma_{\rm crit}} \qquad \mbox{if} \qquad \sigma<\sigma_{\rm crit}\,,
\label{eq:f}
\end{equation}
and $f=1$ otherwise. Here the proportionality coefficient
(``critical'' cross section~\cite{Press:1985ug}) depends on the
neutron star parameters,
\[
\sigma_{\rm crit}\sim R_*^2m_n/M_* \sim 10^{-45}{\rm cm}^2 \;,
\]
where we used ${M_* \sim 1.5\, M_\odot}$ and ${R_{*} \sim 10\;
  \mbox{km}}$ in  the estimate. Notably, $\sigma_{\mathrm{crit}}$ is
comparable to the upper limits on the DM--nucleon cross section
  coming from the direct detection experiments
\cite{LUX:2017ree, XENON:2018voc, PandaX-II:2018woa, XENON:2019rxp,
  PICO:2019vsc}.

Since $dN/dt \propto m^{-1}$, the accumulation rate of dark {\em mass}
does not directly depend on~$m$. Integrating
Eq.~(\ref{eq:capture_gen}), we obtain the total DM mass inside
the 10~Gyr old neutron star, 
\begin{eqnarray}
\text{MW}:~ M_{\rm tot} &\sim& 7\times 10^{42}\,  {\rm GeV} f
\sim 7\times 10^{-15} M_\odot f,
\label{eq:total_capturedMW} \\
\text{dwarf}:~ M_{\rm tot} &\sim& 8\times 10^{46}
\,{\rm GeV} f\sim 7\times 10^{-11} M_\odot f\,.
\label{eq:total_capturedDwarf}
\end{eqnarray}
The above two options differ by  the parameters of the ambient DM
distribution: $\bar v\approx 220$~km/s, ${\rho_{\rm DM} \approx 
0.3}$~GeV/cm$^3$ in the Milky Way (MW) and ${\bar v \sim 7}$~km/s,
${\rho_{\rm   DM} \sim  100}$~GeV/cm$^3$ in the densest dwarf
galaxies~\cite{Strigari:2006rd,Strigari:2007at}. Besides, taking $f=1$ turns
the above
estimates into model--independent upper limits on the total
DM mass that can be captured by the neutron star in the respective
environments.

Another possibility is a gravitational trapping of the DM during
formation of the neutron star~\cite{Capela:2012jz,
  Capela:2014ita}. The latter objects are created
in the supernova collapses of ordinary stars with
masses above $\sim 9M_\odot$, which, in turn, are born
  in giant molecular
clouds. Originally, as the baryonic gas contracts adiabatically into
a proca star, a low--velocity fraction of the ambient DM gets trapped
by its gravitational well. Eventually, this DM ends up in the center
of a heavy progenitor star; estimates of~\cite{Capela:2012jz,
  Capela:2014ita} show that the captured DM mass is close to
Eqs.~(\ref{eq:total_capturedMW}), (\ref{eq:total_capturedDwarf})
within an order of magnitude. Later, a fraction of the 
star DM is inherited by the neutron star in the course of supernova
collapse. That last process has never been considered in the
literature, but our rough estimate suggests that the respective
suppression of $M_{\rm tot}$ lies between~$1$ and~$10^{-3}$.

In the rest of this paper we will use
Eqs.~(\ref{eq:total_capturedMW}), (\ref{eq:total_capturedDwarf}) and
ignore the second, ``gravitational capture'' mechanism despite the fact
that it seems less sensitive to non--gravitational DM interactions. Indeed,
that additional mechanism cannot dramatically change the amount of
captured
DM even in the best case. Besides, it depends on the multi--stage neutron
star formation which is subject to large astrophysical  
uncertainties. Finally, it does in fact rely on the DM--DM and
DM--neutron interactions, albeit in a subtle and indirect way: the 
non--gravitational couplings are needed to thermalize the DM particles
inside the progenitor star and mix them in the phase space, or none
would lose enough energy to get into the neutron star. 


\subsection{Thermalization and condensation}
\label{sec:thermalization}
Once gravitationally bound, the DM particles settle on star--crossing
orbits and continue to lose energies in repeating collisions with
neutrons until a thermal equilibrium is reached. Their kinetic
energies reduce to the neutron star temperature 
${T\sim 10^5}$~K and their orbits shrink to the ``thermal'' radius, 
\begin{equation}
r_{\rm th} \approx \left({9T\over 8\pi G \rho_c m}\right)^{1/2} 
\sim 20\,  {\rm cm} \left( {100\, {\rm GeV}\over m}\right)^{1/2}\,,
\label{eq:r_therm}
\end{equation}
where we substituted the density ${\rho_c \sim 10^{15}}$~g/cm$^3$
of
the neutron star core. The particles steadily arrive into the
central DM cloud during the entire lifetime of the neutron star with
  the rate (\ref{eq:capture_gen}). 

The  characteristic time of this thermalization process was
estimated in Refs.~\cite{Goldman:1989nd, Garani:2020wge}. For almost
any model within our scenario, it is much shorter than the age
of the Universe ~\cite{Goldman:1989nd}. Indeed, we had already
mentioned that BH formation requires a large number of DM
particles inside the neutron star~\cite{Bell:2013xk}. To capture all
of them, one usually assumes the largest possible DM--neutron cross
section ${\sigma\sim\sigma_{\mathrm{crit}}}$, and  even that may
be insufficient. With these interactions, the DM particles equilibrate
quickly, since every star crossing in the beginning of the 
process leads to scattering. On the other hand, we will be able to
consider very small DM--neutron cross sections once the new mechanism
for reducing the required multiplicity to Eq.~(\ref{eq:bosons}) is
invoked. In that case a necessity to thermalize the DM imposes the
strongest constraint on its interactions with neutrons. 

As the DM particles continue to accumulate, the multiplicity $N$ of
the central cloud grows at a fixed radius~(\ref{eq:r_therm}). Eventually,
the mean distance between the particles $r_{\mathrm{th}} N^{-1/3}$ drops below
the size $(mT)^{-1/2}$ of their wave functions. This happens at
\begin{equation}
\mathrm{BEC:} \;\;\;\; N  \gtrsim 0.2 \; T^3 M_{pl}^3
\,\rho_c^{-3/2} \sim 10^{36}\,,
\label{eq:condensate}
\end{equation}
i.e.\ significantly before the maximal number of
particles~(\ref{eq:total_capturedMW}), (\ref{eq:total_capturedDwarf})  
is reached.  At this point the particle wave functions start to overlap
and Bose--Einstein condensate forms in the thermal
cloud~\cite{Semikoz:1994zp, Khlebnikov:1999qy, Levkov:2018kau,   
  Chen:2021oot}. Since then, most
of the thermalized DM particles occupy the
lowest level in the combined DM and neutron star gravitational
potential, with the thermal energy being carried by a  few remaining 
particles.  

Thanks to large occupation numbers~(\ref{eq:condensate}), the
Bose--Einstein condensate at the lowest level can  be described by
a classical DM field $\phi(x)$. From the very beginning it 
forms a non-rotating~\cite{Dmitriev:2021utv} self--bound
soliton which is almost detached from the 
neutron star surroundings. Indeed, even at
multiplicity~\eqref{eq:condensate} the gravitational field 
of this object can be estimated to exceed the neutron star's, and with time
the soliton mass grows. We will call this soliton a Bose
star~\cite{Ruffini:1969qy, Tkachev:1986tr} or a
Q--ball~\cite{Friedberg:1976me, Coleman:1985ki, Lee:1991ax}  if it
is mostly bound by self--gravity or by attractive self--interactions,
respectively. In either case, the properties of the soliton can be
determined by solving the stationary classical field equations, where
a single free parameter is the total number of accumulated DM
particles~$N$.

With the soliton formation the final, model--dependent and nonlinear,
stage of DM evolution begins. As $N$ increases from
Eq.~(\ref{eq:condensate}) to the maximal values in
Eqs.~(\ref{eq:total_capturedMW}), (\ref{eq:total_capturedDwarf}), the
soliton becomes heavier. It collapses gravitationally if the condition
for collapse~--- the hoop conjecture~--- gets satisfied prior to
accumulating the maximal amount of DM. If, to the contrary, the
soliton size exceeds its Schwarzschild radius even for the
largest $N$ in Eq.~(\ref{eq:total_capturedDwarf}), the black hole does
not form. 

Let us finish Section by commenting on DM thermalization in white
  dwarfs~--- second to best compact objects accumulating DM. Their 
  cores are  significantly  more dilute, ${\rho_c \lesssim 10^8 \,
  \mbox{g}/\mbox{cm}^3}$, and have higher typical temperatures
  ${T \gtrsim 10^6 \, \mbox{K}}$ as compared to the neutron stars,
  see~\cite{Chabrier_2000}. Thus, Bose--Einstein condensation of dark
  matter particles in their centers would require larger DM multiplicity 
   $N\gtrsim 10^{49}$, see Eq.~(\ref{eq:condensate}). But in fact, the
  DM inside the thermal radius becomes self--gravitating before that,
  at ${N\gtrsim {10^{46}} \, (m/100\, \mbox{GeV})^{-5/2}}$. It
  collapses gravithermally and forms a compact object where the
  subsequent DM cooling and (possibly) Bose--Einstein condensation
  occur. This process deserves a separate study,
  which is outside of the scope of this paper.


\section{Selection rules}
\label{sec:grow-bose-einst}

\subsection{Optimizing the DM model}
\label{sec:selection-rule-dark}
As we stressed in the Introduction, interactions generically
detain the gravitational collapse until the number of DM particles
  becomes parametrically larger than in Eq.~(\ref{eq:bosons}), and
this is often too large to be accumulated by the neutron star
  during the Universe's lifetime. To gain a more quantitative
understanding of the numbers, we start with the simplest scalar DM
described by a single complex field,
\begin{equation}
  \label{eq:54}
  {\cal L} = |\nabla_\mu \phi|^2 - V(|\phi|)\,.
\end{equation}
We will try to optimize its scalar potential $V$ in a way that
minimizes the dark matter multiplicity required for collapse. We
consider  minimal coupling to gravity and ignore interactions with
the visible matter~--- those will be added later. Importantly, we
also assume that the  typical scale(s) $\Lambda$ in the potential $V$
are essentially sub--Planckian,
\begin{equation}
  \label{eq:50}
  \Lambda \ll  M_{pl}\,.
\end{equation}
This separates our model from the effects of quantum gravity.

Suppose a compact, stationary, and stable solitonic configuration of the scalar
field~--- a Q--ball or a Bose star~--- is formed inside the neutron star. Let
us compute its mass $M$ and radius $R$ as functions of the global
charge~$N$. By conservation law, the latter quantity counts the number of dark matter
particles appended to the Bose--Einstein condensate. We
recall, first, that a generic stationary solution in the
model~(\ref{eq:54}) has the form,
\begin{equation} 
  \label{eq:53}
  \phi = \varphi(\bm x) \; \mathrm{e}^{-i\omega t}\,.
\end{equation}
Here we introduced the real--valued soliton profile
$\varphi(\bm{x})$ and the energy $\omega$ of particles inside it. Second,
the gravitational field of the  soliton is expected to be small,
$g_{\mu\nu} \approx \eta_{\mu\nu}$, except for the critical case when
gravitational collapse is about to happen. In this case
the flat--space Noether charge can be used:
\begin{equation}
  \label{eq:8}
  N \approx i \int d^3 \bm{x} \;\left(\phi^* \partial_0 \phi - \phi
  \partial_0 \phi^*\right) \sim \omega \varphi_0^2 R^3\,, 
\end{equation}
where the last equality is a crude estimate in terms of the
soliton size $R$ and the field $\varphi_0\sim \varphi(0)$ in its
center. Since every particle inside the soliton has energy~$\omega$,
the total mass of this object is of order~${M \sim \omega
  N}$. 

\begin{figure}
  \centerline{\includegraphics{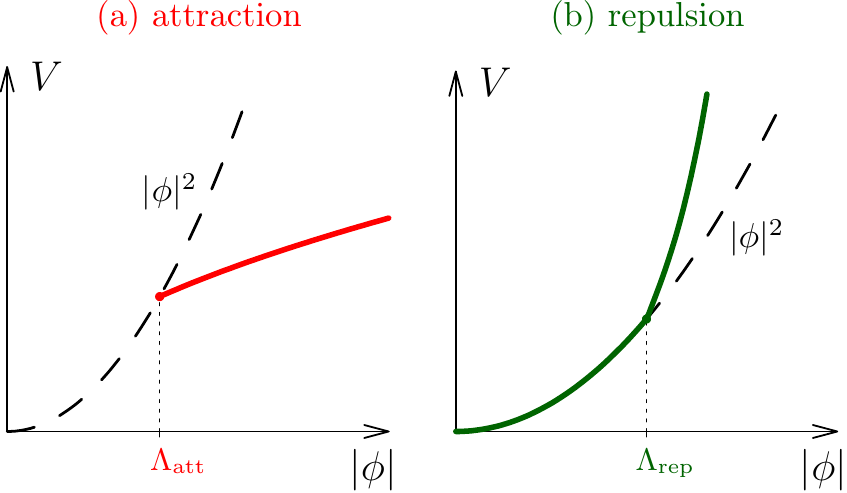}}
  \caption{Possible forms of the scalar potential.}
  \label{fig:V}
\end{figure}

Recall, however, that the soliton parameters $\varphi_0$ and $R$ are related
by the field equations which involve the scalar
potential~$V(\varphi_0)$. We therefore consider two options. First,
  one can assume that the potential grows 
almost quadratically at weak fields, $V \sim   m^2|\phi|^2$, and
then flattens out beyond some scale $\Lambda_{\mathrm{att}}$ like
$V \propto |\phi|^\alpha$ with $\alpha<2$, see
Fig.~\ref{fig:V}(a). This  case corresponds to particle attraction inside
the solitonic core, as the energy per unit charge is smaller than
the particle mass.  At strong fields we approximate,
\begin{equation} 
  \label{eq:57}
  V \approx m^2 |\phi|^{\alpha} \; \Lambda_{\mathrm{att}}^{2-\alpha}  \qquad
  \mbox{at}\qquad |\phi| \gtrsim  \Lambda_{\mathrm{att}} \,,
\end{equation}
where $0<\alpha<2$. It is precisely the scalar
self--interaction that holds the soliton~--- the Coleman's
Q--ball~\cite{Coleman:1985ki}~--- together, since gravity is
weaker: $\Lambda_{\mathrm{att}} \ll  M_{pl}$ according to 
Eq.~\eqref{eq:50}. In the field equation, the self--attraction is 
balanced  by the kinetic pressure: $\partial_i^2 \phi \sim V' \sim
\omega^2 \phi$, or
\begin{equation}
  \label{eq:58}
  \varphi_0 / R^2 \sim m^2 \varphi_0^{\alpha-1}\;
  \Lambda_{\mathrm{att}}^{2- \alpha} \sim \omega^2 \varphi_0 \,,
\end{equation}
see Appendix~\ref{sec:bose-star-conf} for details. This gives
$\omega \sim R^{-1}$,
\begin{equation}
  \label{eq:59}
  N \sim (R\Lambda_{\mathrm{att}})^2 (Rm)^{4/(2 - \alpha)}\;,\;\;\;\;
  \mbox{and} \;\;\;\;
  M \sim N/R \,,
\end{equation}
where we parametrize the soliton configurations with their sizes
$R$.

The soliton collapses gravitationally when its 
compactness $R_g/R=2M/(M_{pl}^2 R)$ becomes of
order one, i.e.\ at masses above critical, 
\begin{equation}
  \label{eq:56}
M_{cr} \approx R M_{pl}^2/2 \sim \omega N_{cr}\,.
\end{equation}
Substituting this relation into Eq.~(\ref{eq:59}), we obtain the
number of DM particles needed for collapse, 
\begin{equation}
  \label{eq:60}
  N_{cr} \sim \frac{M_{pl}^2}{m^2}\,
  \left(\frac{M_{pl}}{\Lambda_{\mathrm{att}}}\right)^{2 - \alpha} \quad
  \mbox{for}  \;\; 0 < \alpha < 2\,.
\end{equation}
At the critical point, the field inside the soliton is Planckian: $\varphi_{0,\,
  cr} \sim M_{pl}$, cf.\ Eqs.~(\ref{eq:58}), (\ref{eq:59}), and \eqref{eq:60}.

Notably, the critical multiplicity \eqref{eq:60} is minimal at ${\alpha
= 2}$ when the ``free bosonic'' expression~(\ref{eq:bosons}) is
recovered. The other values of $\alpha<2$ are less
advantageous, as $\Lambda_{\mathrm{att}}$ is
parametrically below the Planck scale. One obtains almost the
``fermionic'' multiplicity (\ref{eq:fermions}) in the 
case  $V \propto |\phi|$ and even larger $N_{cr}$
for the $\phi$--independent potential. Thus, contrary to naive
expectations particle attraction obstructs 
collapse, the reason being that the energy per particle $\omega$
becomes much smaller than $m$.

One can push the above ``attractive'' option to the extreme assuming
that the scalar potential $V(\phi)$ decreases at strong fields,
e.g.\ $V\propto -|\phi|^\alpha$. However, that would destabilize
the soliton making its field evolve towards lower~$V$,
cf.~\cite{Zakharov12,Levkov:2016rkk}. With no new positive terms
in the potential to stop the process, the region with $V(\phi)< 0$
would be eventually reached~\cite{Panin:2018uoy}; then the soliton
turns into an expanding and Universe--destroying bubble of true 
vacuum~\cite{Coleman:1978ae}. Even if the bubble can be somehow forced 
to collapse gravitationally, the region in its center 
still breaks the positivity conditions, so that a naked
singularity may appear instead of a black hole. If, alternatively, the
potential  starts growing at larger fields, again, the field 
stops rolling at that point, thus bringing us back to the two options in 
Fig.~\ref{fig:V}. 

Equation (\ref{eq:60}) hints at the possibility that the second
option of repulsive self--interactions with $\alpha > 2$ may be
more interesting, see Fig.~\ref{fig:V}(b). In this case the only
attractive force is gravity. The respective soliton is called a Bose
star~\cite{Ruffini:1969qy, Tkachev:1986tr}, since it is bound by
  gravitational attraction compensating interaction pressure of Bose
particles inside it. Note that the repulsive energy gives a subdominant 
contribution to the mass of the subcritical object because its
opponent~--- the  gravitational  energy~--- remains small until the
rim of collapse. We therefore keep two terms in the potential,
\begin{equation}
  \label{eq:61}
  V \approx m^2 |\phi|^2 + m^2|\phi|^{\alpha} \;
  \Lambda_{\mathrm{rep}}^{2-\alpha}   \,, 
\end{equation}
where now $\alpha > 2$, the field $\phi$ is arbitrary, and
$\Lambda_{\mathrm{rep}}$ satisfies Eq.~(\ref{eq:50}). Performing the
estimates similar to the ones before~\cite{Ho:1999hs}, we find out that the 
  Bose star collapses gravitationally only at $\alpha > 8/3$, see
  Appendix~\ref{sec:bose-star-conf} for
  details. In this case the soliton field indeed gets stuck in the
region of subdominant self--interactions ${\varphi_0 \lesssim
  \Lambda_{\mathrm{rep}}}$ until the collapse, at which point
$\varphi_{0,\, cr} \sim \Lambda_{\mathrm{rep}}$ and the soliton
charge equals
\begin{equation}
  \label{eq:62}
  N_{cr} \sim \frac{M_{pl}^3}{\Lambda_{\mathrm{rep}} m^2} \qquad
  \mbox{for $\alpha > 8/3$}\,.
\end{equation}
Notably, this critical multiplicity is independent of~$\alpha$,
see also Refs.~\cite{Colpi:1986ye} and~\cite{Ho:1999hs}. At
$\Lambda_{rep} \sim m$ the 
expression~\eqref{eq:62} reproduces the fermionic
result~(\ref{eq:fermions}).  The
way to decrease the multiplicity  is to increase
$\Lambda_{\mathrm{rep}}$ suppressing the self--repulsion. At
$\Lambda_{\mathrm{rep}}\sim M_{pl}$ that force is as weak as gravity,
the field inside the critical  Bose star is Planckian, and we obtain
the ``free bosonic'' formula~(\ref{eq:bosons}),  again.

The remaining region $2 < \alpha < 8/3$ is worst of them all,
  since in that case the Bose--Einstein condensate does not clump
  under self--gravity. Indeed, an estimate of 
  Appendix~\ref{sec:bose-star-conf} shows that the 
  self--repulsion $|\phi|^\alpha$ with $\alpha$ in this range is stronger than gravity
   at large 
  distances. It  makes the DM condensate spread over the entire volume  
  available inside the neutron star gravitational field.

  We conclude that the critical particle number $N_{cr}$  is smallest in the restricted class of models with
long, Planckian--size valleys ${|\phi| \lesssim M_{pl}}$ and almost
quadratic potentials $V\propto |\phi|^2$ at their bottoms. Any
interaction impedes collapse 
and sharply increases the critical multiplicity. In particular,
self--repulsion becomes strong, almost equivalent to the Fermi pressure at
large fields~--- hence the ``fermionic''  result
(\ref{eq:62}). Self--attraction does not help either: it provides
negative binding energy and lowers the soliton mass which is bad
for collapse. The respective critical particle number 
is also parametrically larger, see Eq.~(\ref{eq:60}). 

In the optimal model with exactly quadratic potential, the kinetic
pressure inside the Bose star is compensated by the gravitational
attraction~\cite{Kaup:1968zz, Ruffini:1969qy, Schunck:2003kk}. The
respective object becomes gravitationally unstable at $\varphi_{0,\,
  cr} \sim M_{pl}$. It has the critical multiplicity~\cite{Kaup:1968zz}
\begin{equation}
  \label{eq:30}
  N_{cr} \approx 0.653 \, M_{pl}^2/m^2 \qquad \mbox{for $\alpha=2$}\,.
\end{equation}
This result agrees with the estimate in
Eq.~(\ref{eq:bosons}). Nonrelativistic approximation remains valid
during the most part of the Bose star growth and gets  marginally
broken at the rim of collapse.

\subsection{Obstruction by quantum corrections}
\label{sec:selection-rules}
How far the DM model can deviate from the free bosonic
theory? To find out, we require that the critical multiplicity
for collapse $N_{cr}$ does not exceed the maximal
amount~(\ref{eq:total_capturedDwarf}) of captured DM:~${mN_{cr}
  \lesssim M_{\rm tot}^{\mathrm{dwarf}}}$. 

In the attractive case, this condition bounds
from below the scale
$|\phi| \sim \Lambda_{\mathrm{att}}$ at which the scalar potential
flattens out in Fig.~\ref{fig:V}(a):
\begin{equation}
  \label{eq:67}
  \Lambda_{\mathrm{att}} \gtrsim M_{pl} \left( 
  \frac{2 \times 10^{-9}\; {\rm GeV}}{f\; m}\right)^{1/(2 - \alpha)}
  \!\!\!\! ,
\end{equation}
see Eqs.~(\ref{eq:total_capturedDwarf}), (\ref{eq:60}) and recall
that $0< \alpha < 2$. Typically, $\Lambda_{\mathrm{att}}$ is very
large. Indeed, since the critical Q--ball has $\varphi_0
\sim M_{pl}$, the model with flat potential should be trustable,  
i.e.\ renormalizable and weakly coupled, all the way up to the
Planckian scale. The only manifestly renormalizable flat potential is $V
=\mathrm{const}$, it appears~\cite{Levkov:2017paj} e.g. in the 
celebrated Friedberg--Lee--Sirlin model~\cite{Friedberg:1976me}. In
this case 
\begin{equation}
  \label{eq:64}
  \Lambda_{\mathrm{att}} \gtrsim 5\times 10^{13}\; \mbox{GeV} \;
  (mf/100 \, \mathrm{GeV})^{-1/2}\;, \;\;\alpha = 0\,.
\end{equation}
Soon we will see that large scales (\ref{eq:67}), (\ref{eq:64}) are
problematic because multiloop corrections to the scalar potential
become relevant at $\varphi_0 \ll \Lambda_{\mathrm{att}}$. They
generate interaction  pressure and prevent collapse.

The scale $\Lambda_{\mathrm{att}}$ can be substantially lowered in
a specific class of renormalizable multifield models where the potentials 
at the bottoms of curved valleys have ${\alpha \approx 2}$. We will
consider this option in Sec.~\ref{sec:models-with-bended}.

In the opposite, self--repulsive case  the inequality
$mN_{cr} \lesssim M_{\rm tot}^{\mathrm{dwarf}}$ also provides a large
scale
\begin{equation}
  \label{eq:65}
  \Lambda_{\mathrm{rep}} \gtrsim 2\times 10^8\; \mbox{GeV}\; 
(mf/100 \, \mathrm{GeV})^{-1}\,,
\end{equation}
following from Eqs.~(\ref{eq:total_capturedDwarf}) and (\ref{eq:62}).
This condition strongly suppresses all
repulsive self--interactions at fields ${|\phi| \lesssim
  \Lambda_{\mathrm{rep}}}$, see Eq.~(\ref{eq:61}). For  example, in
the~$\lambda_4 |\phi|^4/4$ case,
\begin{equation}
  \label{eq:66}
  \lambda_4 \equiv (2m/\Lambda_{\mathrm{rep}})^2 \lesssim  
7\times 10^{-13} f^2
  (m/100 \, \mathrm{GeV})^4\,.
\end{equation}
It is worth remarking that the renormalizability of the repulsive
potential is not required, since the Bose star field ${\varphi_{0} \lesssim
  \Lambda_{\mathrm{rep}}}$ remains parametrically below the natural
cutoff of the theory~(\ref{eq:61}) prior to collapse. 

Now, recall that our dark matter should interact with the visible
sector, and strongly enough, in order to be captured by
   the neutron
star. The respective couplings should be renormalizable and stay under
control at strong fields~---  hence, their forms are 
constrained by the Standard Model symmetries.  Let us demonstrate that,
generically, loop corrections from these interactions break the desired 
properties of the dark matter potential: lift its flat parts  and
generate unacceptably large repulsive vertices, as has been
first pointed out in~\cite{Bell:2013xk}.

\begin{figure}
  \centerline{\includegraphics{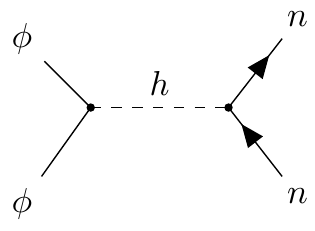}}
  \caption{Scattering $\phi n \to \phi n $.}
  \label{fig:dia}
\end{figure}

Couple, e.g., the field $\phi$ to the Higgs doublet $H(x)$ by
  deforming the potential of the latter to
\begin{equation}
  \label{eq:4}
V =  \lambda_H \left(H^\dag H -
\frac{v^2}{2} - \frac{y|\phi|^2}{2 \lambda_H}\right)^2 + m^2 |\phi|^2\,,
\end{equation}
cf.~\cite{Shaposhnikov:2006xi, Bezrukov:2009yw}. Here $v \approx
  246\, \mbox{GeV}$ and ${\lambda_H \approx 0.13}$ are the
standard Higgs parameters, and the new constant $y$ regulates
its interactions with the dark sector. Notably, the model is weakly 
coupled at  ${y\ll (4\pi\lambda_H)^{1/2} \sim 1}$.

The physical Higgs field $h(x)$ is defined as ${H^\dag H =
    (v+h)^2/2}$. It interacts with neutrons $n$
via the effective Yukawa vertex ${V_{hnn}   \approx 2m_n h
  \bar{n}n/(9v)}$ which is known up to light quark
contributions giving a factor $1\div  2$~\cite{Shifman:1978zn,
  Cirelli:2013ufw}; $m_n \approx \mbox{GeV}$ is the neutron
  mass. This means that the dark matter also scatters off 
neutrons. The respective diagram is shown in Fig.~\ref{fig:dia} and
the cross section at nonrelativistic momenta equals\footnote{Hadronic
  form factors and strong nucleon interactions may suppress this cross
  section by an additional factor of up to $10^{-3}$,
    see~\cite{Bell:2020obw, Anzuini:2021lnv}. However, that would
  only sharpen the arguments of this section.}
\begin{equation}
  \label{eq:38}
  \sigma = \frac{y^2 m_n^4}{81 \pi m_H^4 m^2}\,,
\end{equation}
where $m_{H} = v (2\lambda_H)^{1/2} \approx
125\, \mbox{GeV}$.
Importantly, the scattering probability
${f   \equiv \sigma / \sigma_{\mathrm{crit}} <   1}$ should be
large enough to capture the $\phi$--particles inside the neutron star,
see the constraints (\ref{eq:67}), (\ref{eq:65}). Thus, the  coupling
$y$ is large.

On the other hand, the same coupling \eqref{eq:4} generates DM
  pressure at strong fields. In fact, we have already tuned this
  potential to cancel  DM self--interactions
  at the classical
  level. Namely, at large fixed $\varphi_0 = |\phi|$ the Higgs field
  adjusts itself to minimize the first term in
  Eq.~(\ref{eq:4}),
\begin{equation}
  \label{eq:3}
  H^{\dag}H \equiv  \frac12 (h+v)^2 = \frac{v^2}{2} +
  \frac{y|\phi|^2}{2\lambda_H}\,.
\end{equation}
As a consequence, the tree potential is quadratic at the bottom of this potential
valley, ${V\approx m^2 |\phi|^2}$. Nevertheless, the self--coupling
reappears, again, once loop contributions from the visible matter are 
included. As an illustration, let us first ignore all fields except
for the Higgs boson and $\phi$ itself. Then their one--loop
effective potential~\cite{Weinberg:1996kr} at large $|\phi|
\propto h$ along the valley~\eqref{eq:3} equals,
 \begin{equation}
  \label{eq:37}
  \Delta V_{\mathrm{1-loop}} \approx \lambda_{4,\, \mathrm{eff}}\, |\phi|^4/4\,.
\end{equation}
where  $\lambda_{4,\, \mathrm{eff}} = y^2 L/ (2\pi^2) + O(y^3 /
\lambda_H)$ is a correction to the $|\phi|^4$ self--coupling, ${L =\ln ( |\phi| /
  \Lambda_{\mathrm{ren}})}$, and ${\Lambda_{\mathrm{ren}} \sim m}$ is
  a renormalization scale. Note that a proper calculation of the
effective potential at strong fields includes renormalization group
resummation of the leading logs~\cite{Bednyakov:2012en,
  Chetyrkin:2013wya}. However, that usually introduces an
order--one factor in $\lambda_{4,\, \mathrm{eff}}$ which does not affect
our estimates.

We see that Eq.~(\ref{eq:37}) brings in  the $|\phi|^4$ repulsion
even if it was absent before. Moreover, since $\lambda_{4,\,
 \mathrm{eff}}$ logarithmically depends on the 
field, the new contribution cannot be canceled by any renormalizable
counter-terms even if an arbitrarily precise fine--tuning is
allowed. An obvious way out is to make the one--loop repulsion small,
so that it does not preclude black hole formation. Requiring
$\lambda_{4,\,   \mathrm{eff}}$ to satisfy Eq.~(\ref{eq:66}) with
$\sigma$ given by Eq.~(\ref{eq:38}), we obtain the inequality
\begin{equation}
  \label{eq:25}
  y \gtrsim 400 \, L^{1/2} > 400\,,
\end{equation}
which cannot be satisfied in a weakly coupled model with $y\ll
  1$. Moreover, even this unacceptable lower limit can be
  achieved only if $m \gtrsim {\rm PeV}$.
We conclude that either the coupling constant $y$ is too small for
  accumulating the required amount of $\phi$--particles, or
the interaction pressure caused by the same constant prevents the  
$\phi$--condensate from collapsing.

As an alternative, one may try to couple $\phi$ to fermions that 
generate negative terms in the effective potential. In the model
(\ref{eq:4}) this  amounts to recalling that every massive
Standard Model field
 gives a potential to Higgs and therefore
produces a vertex $|\phi|^4 \propto h^4$ along the
valley~(\ref{eq:3}). The leading contribution is negative and comes
from the top quark Yukawa coupling $y_t\approx  1$. We obtain $\delta
\lambda_{4,\, \mathrm{eff}} = - 3y^2y_t^2 L/(8\pi^2 \lambda_H^2)$. This
definitely destabilizes the   valley~\cite{Buttazzo:2013uya,
  Bednyakov:2015sca} unless even larger positive $|\phi|^4$ terms are
introduced, which,  however, returns us to the no--go
estimates given above.

Somewhat more elegantly, one may organize a (partial) cancellation
between the fermionic and bosonic loops. But that would mean upgrading
the Standard Model to (N)MSSM. In that case the inequality
  $mN_{cr} \lesssim M_{\rm tot}^{\mathrm{dwarf}}$ imposes strong
constraints on the
supersymmetry--breaking
operators that detune the cancellation~\cite{Bell:2013xk}. In
Sec.~\ref{sec:models-with-bended} we consider more economic and
general possibility. 

To sum up, loop corrections from the visible sector are
dangerous, generic, and cannot be avoided. Thus, we need a
special mechanism to tame them. 

\subsection{Requirements for the DM model}
\label{sec:requirements}
Let us summarize the properties of  dark matter model needed
for black hole formation inside the neutron star.

(a) The scalar potential of the model should include a long valley
parametrized by the complex scalar $\phi$. The valley should extend
to large fields: to $|\phi| \sim M_{pl}$ in the case of
attractive self--interactions (\ref{eq:57}) or, in the
repulsive case~(\ref{eq:61}), at least to the scale ${|\phi| \sim  
\Lambda_{\mathrm{rep}}}$ given in Eq.~(\ref{eq:65}). The model
should remain weakly coupled at these fields i.e.\ be
renormalizable or have a sufficiently high cutoff.

(b) The potential should be
almost quadratic along the valley,  ${V \approx  m^2
  |\phi|^2}$. All its repulsive terms should be
suppressed at least by the scale $\Lambda_{\mathrm{rep}}$ in
Eqs.~(\ref{eq:61}), \eqref{eq:65}. The potential
may become flatter than quadratic 
(attractive) inside finite $\phi$ intervals, but that
should  not ruin its renormalizability  or
destabilize the vacuum. If the potential becomes attractive
asymptotically at ${|\phi| \gtrsim \Lambda_{\mathrm{att}}}$, like in
Eq.~(\ref{eq:57}), the scale $\Lambda_{\mathrm{att}}$ should be
sufficiently high, see Eq.~(\ref{eq:67}).

(c) Quantum corrections from the dark matter and visible sectors
should not destabilize the valley or create effective interactions breaking
the condition (b).

We have already demonstrated that the condition (c) is usually
violated by dark matter interactions  with the visible 
sector in one--field DM models. We now turn to models where this can
be avoided.

\section{Models with bended valleys}
\label{sec:models-with-bended}

\subsection{The mechanism}
\label{sec:main-mechanism}
Start from the arbitrary model for the DM field $\phi$ and add the
  second complex scalar $\chi$ in such a way that $\phi$ and
$\chi$ have global charges 1 and~2, respectively.   
 The coupling between the two is then chosen in a specific
renormalizable and positive--definite form,
\begin{equation}
  \label{eq:31}
  V = \lambda \left|\phi^2 - \Lambda \chi\right|^2 
    + m^2 |\phi|^2 + \lambda'\, |\phi|^4/4\,,
\end{equation}
where the last two terms represent the original $\phi$ potential.
This model is invariant under phase rotations ${\phi \to
  \mathrm{e}^{i\theta}\phi}$ and $\chi \to\mathrm{e}^{2i\theta}\chi$.
In the vacuum $\phi = \chi =  0$ all fields are massive: ${m_\phi =
  m}$ and ${m_{\chi} =   \Lambda\lambda^{1/2}}$. We will assume that
the first term in the potential is the largest:  
\begin{equation}
  \label{eq:1}
  m \ll \Lambda \lambda^{1/2} \qquad \mbox{and} \qquad \lambda' \ll
  \lambda\,.
\end{equation}
This means, in particular, that the $\chi$--particles are heavy and do
not change cosmology.

The trick of the new model is to make the potential valley {\it bend}
in the $\phi$---$\chi$ space. Indeed, the low--energy field configurations
are expected to minimize the largest term in Eq.~(\ref{eq:31}),
i.e.\ satisfy
\begin{equation} 
  \label{eq:29}
  \chi = \phi^2/\Lambda\,.
\end{equation}
The other terms of $V$ create a small potential
along this valley. The same is true, in particular, for the stationary
nonrelativistic soliton, 
\begin{equation} 
  \label{eq:32}
  \phi = \varphi(\boldsymbol{x}) \, \mathrm{e}^{-i\omega t}\;,\qquad  
  \chi = \tilde{\chi}(\boldsymbol{x}) \, \mathrm{e}^{-2i\omega t}\,,
\end{equation}
which has  real profiles $\varphi$, $\tilde{\chi}$ satisfying
Eq.~(\ref{eq:29}).  

Let us explicitly demonstrate that the valley extends in the $\phi$
direction at small fields, then takes a turn at ${\phi \sim \chi \sim
\Lambda}$ and goes along~$\chi$. To this end we introduce a
combination $\rho(\boldsymbol{x})$ of the solitonic profiles
which has a
canonical kinetic term along the valley: $(\partial_i \rho)^2 =
(\partial_i \varphi)^2 + (\partial_i \tilde{\chi})^2$. Using
Eq.~(\ref{eq:29}) and integrating, we find 
\begin{equation}
  \label{eq:5}
  \rho(\varphi) = \frac{\varphi}{2}\sqrt{1+
    \frac{4\varphi^2}{\Lambda^2}} + \frac{\Lambda}{4}\, \mathrm{arcsinh}
  \left( \frac{2\varphi}{\Lambda}\right)\,.
\end{equation}
The classical potential $V(\rho)$ at the bottom of the valley is
obtained by inverting Eq.~\eqref{eq:5} and substituting
$\varphi(\rho)$ into the last two terms of Eq.~(\ref{eq:31}). 

At small fields we get $\rho \approx \varphi$; hence, the valley
stretches in the $\phi$ direction, indeed. Taylor series expansion in
$\rho/  \Lambda \ll 1$ then gives a nonlinear valley potential,  
\begin{equation}
  \label{eq:2}
  V(\rho) =  m^2
  \rho^2 + \lambda_\rho \rho^4/4 + O(\rho^6/\Lambda^4)
\end{equation}
with ${\lambda_{\rho}  =  \lambda'  - 16m^2/(3\Lambda^2)}$. We
have already discussed in Sec.~\ref{sec:grow-bose-einst} that the
$\rho^4$ term cannot be large positive, or it would stop the soliton
from growing. Requiring
\begin{equation}
  \label{eq:55}
  \lambda'/4 = \beta (m/\Lambda)^2\;, \qquad \beta \lesssim 1\,,
\end{equation}
we ensure that the effective coupling is attractive at
small fields: ${\lambda_\rho < 0}$. This is not dangerous for vacuum
stability, since the overall potential of our model is explicitly
positive--definite.

The region $\phi \gg \Lambda$ is entirely
different. Expression~(\ref{eq:5}) gives $\rho \approx \varphi^2 /
\Lambda$ meaning that the valley runs along $\tilde{\chi} \approx  
\rho$. At the same time, the potential at the bottom of the valley
is quadratic: ${V(\rho) = \mu^2\rho^2 + }(\mbox{smaller
  terms})$, where we denoted
\begin{equation}
  \label{eq:63}
  \mu = m\beta^{1/2} \lesssim m\,,
\end{equation}
and used Eqs.~(\ref{eq:31}),~(\ref{eq:55}). So, quartic
self--interaction in this part of the valley is absent, and the
effective $\rho$ mass is smaller than $m$.

Now we see how the bended valley~(\ref{eq:29}) works. Originally,
the model~(\ref{eq:31}) includes the term $\lambda'|\phi|^4$
creating pressure. The same term, however, reduces to a mass
$|\phi|^4\propto |\chi|^2$ once the valley takes a 
turn. Moreover, this property is valid even at the quantum level.

Indeed, an explicit one--loop calculation~\cite{Weinberg:1996kr}
gives  quartic
effective potential at large fields: $\Delta 
  V_{\mathrm{1-loop}} \approx \lambda_{4,\,  \mathrm{eff}} \; 
  |\phi|^4/4 + O(\phi^2)$ with
\begin{equation}
  \label{eq:7}
  \lambda_{4,\, \mathrm{eff}} = \frac{2}{\pi^2} (2\lambda^2 +
  \lambda\lambda' + 5\lambda'^2/32)
  \,\ln\frac{|\phi|}{\Lambda_{\mathrm{ren}}}\;.
\end{equation}
Thus, quantum fluctuations of $\phi$ and $\chi$ somewhat increase
the constant ${\lambda'  
  \to \lambda' +\lambda_{4,\, \mathrm{eff}}}$ in 
front of ${|\phi|^4 \propto |\chi|^2}$, but do not  prevent the Bose
star from growing if Eq.~(\ref{eq:55})
remains satisfied:
\begin{equation}
  \label{eq:71}
  \lambda' + \lambda_{4,\mathrm{eff}} < 4 (m/\Lambda)^2\,.
\end{equation}
The latter inequality is easily met if $\Lambda$ is not too far away from $m$. 

One may wonder, why the dangerous terms $\Delta V = c_1 |\phi \chi|^2
+ c_2 |\chi|^4$ are not generated at the one--loop level. If
present, they would create an undesired pressure inside the
strong--field soliton. However, the auxiliary field $\chi$ enters the
interaction potential in the combination $\Lambda \chi$, where
$\Lambda$ has mass dimension~1. On dimensional grounds, $c_2 \sim
c_1^2  \sim (\Lambda / \Lambda_{\mathrm{cutoff}})^4$, where
$\Lambda_{\mathrm{cutoff}}$ is a cutoff for loops. Thus, the
multiloop diagrams for the dangerous terms converge, the  
constants $c_{1}$, $c_2$ do not depend logarithmically on the  
fields, and one can tune them to zero at $\phi,\, \chi \gg
\Lambda$. Barring this fine--tuning, the quantum corrections are
harmless in the model (\ref{eq:31}).

Let us  visualize the growth of the Bose--Einstein condensate in the
model of this Section. Initially, it forms a Bose star held by the
gravitational forces. This object becomes denser with time 
due to continuous inflow of dark matter particles. At
${\rho \sim m^2/(|\lambda_\rho|  M_{pl}) \ll \Lambda}$ the $\lambda_\rho \rho^4$
self--attraction overcomes the kinetic pressure, and the
Bose star collapses as a bosenova~\cite{Zakharov12,
  Chavanis:2011zi}. This means that its central part starts squeezing
in a particular self--similar fashion~\cite{Zakharov12,
  Levkov:2016rkk} developing strong fields. The squeeze--in halts when
the field in the center reaches ${\rho \sim  \Lambda}$ and the valley
potential stops being attractive, cf.~\cite{Eby:2016cnq, Levkov:2016rkk}.

This is the moment  when a droplet of much denser condensate with $\rho \sim
\chi \gtrsim \Lambda$ appears in the Bose star center. At first, it
has the field strength $\chi \sim \Lambda$. However, the droplet is expected to grow in
density towards $\chi \gg \Lambda$ as more $\chi$--particles join
in. Indeed, once it is there, a  condensation process
$\phi \phi \to \chi$ with energy release $2m - \mu$ becomes
allowed. Even if one assumes that this direct process is ineffective
dynamically, the condensation  should
continue in other, recurrent way. Without direct transmutation
into $\chi$, the $\phi$ particles would create 
another dilute Bose star around the $\chi$--droplet, and that star
would collapse, again, feeding the droplet. In any case all global
charge should be  eventually transported into the dense $\chi$--cloud
and may never leave it due to  large binding energy $2m - \mu$
of the
$\chi$ particles, cf.\ Eq.~(\ref{eq:63}).

Notably, we expect that the central $\chi$--cloud quickly thermalizes
into a dense and non--interacting Bose star with $\chi \gg \Lambda$. Indeed, the
$\chi$--particles effectively scatter with self--coupling $\gtrsim
\lambda'$ at the cloud boundaries. In there, ${\chi \sim \Lambda}$
and hence the  particle number density is huge: $n \sim \mu 
\Lambda^2$. A conservative  estimate constrains the relaxation 
time from the above~\cite{Zakharov-Lvov, Semikoz:1994zp, Levkov:2018kau},
\begin{equation}
  \label{eq:68}
  \tau_{\chi} \sim (\xi \sigma_{\phi \phi} v_\chi n f)^{-1} \ll
  40\, \mbox{yr} \; v_{\chi}^2 \; 
    \frac{(100\, \mathrm{GeV})^3}{\mu \Lambda^2}\,,
\end{equation}
where $\xi \gg (\Lambda / M_{pl})^{2}$ is the time fraction
spent by the particles at the cloud periphery, $\sigma_{\phi\phi}  \gtrsim
\lambda'^2 /(64 \pi \mu^2)$ is their cross section, $f \propto n/(\mu
v_\chi)^3$ is the phase--space density, $v_\chi$ is
velocity, and we used Eqs.~\eqref{eq:55}, \eqref{eq:63}.
Even the most
conservative choices of parameters give small $\tau_\chi$ compared to
the lifetime of the Universe~-- thus, the relaxation is effective, indeed.

Now, we recall that the self--interactions are absent at ${\chi \gg
\Lambda}$. This means that the respective Bose star will grow
pressureless until it collapses gravitationally into a black hole. The
latter event requires the critical charge 
\begin{equation}
  \label{eq:69}
  N_{cr} \approx 1.3 \, M_{pl}^2/(\beta m^2)\,,
\end{equation}
where the numerical coefficient is larger by a factor $2/\beta$ than
in Eq.~(\ref{eq:30}) because $\chi \sim \rho$ has global charge 2
and mass~${\mu = m \beta^{1/2}}$.

We thus constructed a model with almost an optimal critical
multiplicity for the collapse of Bose--Einstein condensate into a
black hole. In the next Section we will see that interaction with
the visible sector and hence DM capture can be added to
this model without disrupting the picture.


\subsection{Adding interactions with the visible sector}
\label{sec:adding-inter-with}
We couple the DM field to the Higgs doublet in the same way
as before~--- by adding the first term in Eq.~(\ref{eq:4}) to the
scalar potential~(\ref{eq:31}). This makes the $\phi$--particles
scatter off neutrons with the cross section $\sigma
\propto y^2$ in Eq.~(\ref{eq:38}).  Importantly, we leave $\chi$ to interact only
with~$\phi$. This maintains the desired properties of the
strong--field Bose stars in our model.

Now, the Higgs field changes along the potential
valley~(\ref{eq:29}) according to Eq.~(\ref{eq:3}). As a
consequence, a nonzero Higgs profile $h(\boldsymbol{x})$ is
generated inside every stationary soliton. A combination of the
three profiles~--- $\varphi$, $\tilde{\chi}$, and $h$~--- with
the canonical kinetic term is
\begin{equation} 
  \label{eq:33}
  \rho(\varphi) = \int_0^\varphi d\varphi' \sqrt{1 +
    \frac{4\varphi'^2}{\Lambda^2} + \frac{y^2 \varphi'^2}{\lambda_H
      (m_H^2 + 2y\varphi'^2)}}\,,
\end{equation}
where the last contribution comes from the Higgs field.

Notably, with interactions (\ref{eq:4}) included, the effect of the
bended valley remains essentially the same. Indeed,
$\rho$ is still proportional to $\varphi$ and 
$\tilde{\chi}$ at $\rho \ll  \Lambda$ and $\rho \gg \Lambda$,
respectively. In the 
weak--field regime $\rho \lesssim \Lambda$ the valley potential is
nonlinear, Eq.~(\ref{eq:2}), and has almost the same quartic constant
$\lambda_\rho$ as before. We
therefore again impose the condition (\ref{eq:55}) to make
$\lambda_\rho$ negative at weak fields. Once the valley turns at $\rho
\gg \Lambda$, the potential $V(\rho)$ becomes quadratic with  mass
(\ref{eq:63}), as guaranteed by the mechanism of the previous
Section. 

All these nice properties remain valid because $\chi$ does not couple
directly to the Standard Model fields and still enters the overall
potential in the combination $\Lambda \chi$ with dimensionful 
$\Lambda$. This makes the nonlinear interactions vanish in the
strong--field region, and they cannot be generated by
loops. We check the latter property by computing the one--loop
effective potential in the three--field model of $\phi$, $\chi$,
and $H$ at the bottom of the valley~\eqref{eq:3},
  \eqref{eq:29}. The result at large fields is given by Eq.~\eqref{eq:37}, where
  the constant $\lambda_{4,\, \mathrm{eff}}$ equals
  Eq.~\eqref{eq:7} plus the Higgs contribution 
\begin{equation}
  \notag
  \delta \lambda_{4,\, \mathrm{eff}} =  \frac{y^2}{\pi^2 \lambda_H}
  \!\left( \lambda + \frac{\lambda_H}{2} + \frac{y}{2} +
    \frac{3\lambda'}{8} + \frac{y^2}{8\lambda_H}\right)  \ln 
  \frac{|\phi|}{\Lambda_{\mathrm{ren}}}\,.
\end{equation}
Similarly, the other Standard Model fields are expected to
generate $\Delta V_{\mathrm{1-loop}} \propto h^4 \propto |\phi|^4$.
Regardless of the value of $\lambda_{4,\, \mathrm{eff}}$, the loop
  corrections safely change the mass $|\phi|^4 \propto |\chi|^2$
in the strong--field region. The value of $\mu$ then can be
  made positive by adjusting $\lambda'$. For simplicity we 
  assume below that the loop corrections are small compared to~$\lambda'$.

Without the interaction pressure, the strong--field Bose star
collapses gravitationally at almost optimal critical
charge~(\ref{eq:69}). This value cannot exceed the total number of
captured dark matter particles, $M_{\rm tot}^{\mathrm{(dwarf)}}/m >
N_{cr}$. One obtains the inequality 
\begin{equation}
  \label{eq:6}
  y > 10^{-7} \, \left(m \over 100 \, \mbox{GeV}\right)^{1/2} \beta^{-1/2}
\end{equation}
 using Eqs.~\eqref{eq:total_capturedDwarf}, \eqref{eq:38}, and
 (\ref{eq:69}). This constraint is relatively mild and can be easily
   satisfied.

 Indeed, let us show that  a viable parametric region for our
   model can be selected. One starts by specifying $m$ within the WIMP
   mass range and $\beta \lesssim 1$. The coupling constant $y$ is then chosen to
   satisfy Eq.~\eqref{eq:6} together with the constraints from the DM detection
   experiments. Below we will argue that a narrower region
   \begin{equation}
     \label{eq:10}
    2\times 10^{-6} \lesssim y \; \left( m \over 100 \,
    \mbox{GeV}\right)^{-1/2} \lesssim 2\times 10^{-5} \, \beta^{-1/2}
  \end{equation}
   is better for phenomenology. The next parameters are $\lambda$ and
  $\lambda' \ll \lambda$. In the simplest case one takes ${\lambda
    \sim y/\lambda_{H}^{1/2}}$ and $\lambda' \gtrsim \lambda^2$,
     thus suppressing the
  running of $\lambda'$: $\lambda_{4,\,
    \mathrm{eff}}$, $\delta \lambda_{4,\, \mathrm{eff}} \lesssim
  \lambda'$. Finally, the scale $\Lambda$ is given by
  Eq.~(\ref{eq:55}). We conclude that all the constraints can be satisfied if the
 hierarchy 
 \begin{equation}
  \label{eq:72}
\lambda' \sim \lambda^2 \sim y^2 / \lambda_H\;, \qquad \Lambda \sim
  m/\lambda\;, \qquad \beta \lesssim 1
\end{equation}
is valid and $y$ belongs to the interval \eqref{eq:10}. 

\subsection{Generalizations}
\label{sec:generalizations}
So far we relied on fine--tuning to kill the $|\phi\chi|^2$ and
$|\chi|^4$ terms in the potential. If added, these interactions
should be extremely weak~--- say, the operator $\lambda_2'  |\chi|^4/4$ should
satisfy the analog of Eq.~\eqref{eq:66}:
  \begin{equation}
    \notag
    \lambda_2' \lesssim 2\times 10^{-13}\, f^2 \beta^3  \left( m / 100 \, \mbox{GeV}
    \right)^4\,.
  \end{equation}
Can we allow stronger
interactions at $|\chi| \gtrsim \Lambda$?

The answer is yes, if extra auxiliary  scalars are introduced. Namely,
let us add the $|\chi|^4$ term along with the new field $\chi_2$
that has a global charge~4. The new terms in the potential are
\begin{equation} 
  \label{eq:73}
  V_2  = \lambda_2 |\chi^2 - \Lambda_2 \chi_2|^2 + \lambda'_2 |\chi|^4/4\,,
\end{equation}
where $\lambda_2' = 4\beta_2  (\mu/\Lambda_2)^2$  and $\beta_2
\lesssim 1$; cf.\ Eq.~(\ref{eq:55}).  Now, the valley takes
a second turn at ${\chi \sim \chi_2 \sim \Lambda_2}$, before the 
$|\chi|^4$ interaction becomes relevant.

One can continue this procedure and arrive at the ``clockwork--like'' model
with $n$ fields $\chi_i$, a hierarchy of scales $\Lambda \ll
\Lambda_2 \ll \dots \ll \Lambda_n$, masses at strong fields ${\mu_i =
      \mu_{i-1}\beta_i^{1/2}}$, and many coupling constants ${\lambda_i'
    \equiv  (2\mu_{i}/\Lambda_i)^2}$, where $\beta_{i}
\lesssim 1$. The Bose star made of the last field $\chi_n$
collapses gravitationally at the critical charge
\begin{equation}
  \label{eq:9}
  N_{cr} = (M_{pl} / m)^2 \prod_{i=1}^{n} (2/\beta_i)\,.
\end{equation}
If $n\sim O(1)$, this multiplicity is  not exceedingly large.

Of course, the field $\chi$ can be made interacting in other,
old--fashioned ways~--- say, by supersymmetry. To this end one
upgrades this field to the supersymmetric sector with a
flat direction~\cite{Gherghetta:1995dv} which represents the
$|\chi|
\gtrsim \Lambda$ part of the 
valley. The flat direction should be lifted by soft terms giving the
mass to the condensate. Then the superpotential will be
protected from quantum corrections  as long as its
fields couple to the visible matter via the dimensionful constant
$\Lambda$.
  
\section{Conclusions}
\label{sec:conclusions}

We considered formation of black holes with masses~${\approx 1\,
  M_\odot}$ by means of bosonic dark matter collapse inside 
neutron stars. This scenario includes DM capture by the neutron stars,
its thermalization with neutrons, Bose--Einstein condensation, and at
last~--- gravitational collapse into seed black holes which eventually
consume the stars.
The overall process includes two conflicting requirements on the DM
model~\cite{Bell:2013xk}. On the one hand, the DM can be captured only
if it interacts strongly enough with the visible sector. On the other
hand, loop contributions of the same interactions generate DM
  self--couplings and hence 
pressure inside dense DM clouds, impeding  their
collapse. We have found, in agreement with Ref.~\cite{Bell:2013xk}, that the
conflict cannot be resolved by tuning the parameters of
the single--field DM models, neither can it be achieved by
optimization of their  scalar potentials. 

Using the crucial observation that the conflicting requirements
are imposed at different scales, we proposed a mechanism that 
  enables
black hole formation within this scenario. The respective DM models 
are deformed at strong fields, i.e.\ in the regions inaccessible to
the direct experiments. Their scalar potentials include bended valleys 
that go along the dark matter scalar~$\phi$ at 
$|\phi| \lesssim \Lambda$ and then turn in the
direction of a new field~$\chi$. In this case $\chi$
self--pressure can be made small and, notably, can be protected
from all quantum corrections by super--renormalizable couplings to
other fields, supersymmetry, or a  ``clockwork--like'' mechanism. As a
consequence, the growth of the DM cloud inside the neutron star
includes a phase transition $\phi\phi \to \chi$ at a certain density
followed by the gravitational collapse of the pressureless $\chi$
condensate.

In our model formation of a black hole requires a relatively
small number Eq.~\eqref{eq:69} of DM particles which can be accumulated
inside the neutron star even at extremely weak couplings, see e.g.\ 
Eq.~\eqref{eq:6} in the case of interaction~\eqref{eq:4}. However, 
other conditions become relevant  at this point. First, the newborn
seed black hole cannot be too light,  $M_{\mathrm{BH}} \gtrsim
10^{13}\, \mbox{g}$, or it evaporates faster than it accretes
neutrons~\cite{Kouvaris:2011fi}. In our model $M_{\mathrm{BH}} \approx
\mu N_{cr}/2$, so this condition constrains an effective mass $\mu$
of the auxiliary scalar $\chi$ at strong
fields,
\begin{equation} 
  \label{eq:11}
  \mu \equiv m\beta^{1/2} \lesssim 17 \; \mbox{GeV}\,,
\end{equation}
where Eq.~(\ref{eq:69}) was used, $m$ is the mass of the DM
particles, and $\beta \lesssim 1$ parametrizes their
self--interactions via Eq.~\eqref{eq:55}. Taking sufficiently small
$\beta$, one fulfills Eq.~(\ref{eq:11}) in the entire WIMP mass
range.

Second,  thermalization of DM with neutrons is an obligatory
part of our scenario needed to form a dense central cloud. But at weak
couplings the equilibrium may be unachievable even on the cosmological
timescales, since  DM interactions with neutrons are
further suppressed at low energies by Pauli
blocking~\cite{Goldman:1989nd}. For 
scalar vertices, the respective  thermalization time was estimated
in~\cite{Garani:2020wge} as    $\tau_{\mathrm{NS}} \approx 35 \pi^2 /
(12 \sigma m T^2)$, where 
${T\sim 10^{5} \, \mbox{K}}$ is the neutron star temperature and
$\sigma$ is the DM--neutron cross section
in vacuum, Eq.~\eqref{eq:38}. Requiring ${\tau_{\mathrm{NS}}
\lesssim 
10^{10}}$~yr, we  constrain the coupling constant,
\begin{equation}
  \label{eq:12}
  y \gtrsim 2\times 10^{-6} \, \left(m \over 100 \, \mbox{GeV}\right)^{1/2}\,.
\end{equation}
This condition is marginally stronger than Eq.~(\ref{eq:6}) 
needed for capture. 

Third and finally, our mechanism should not be too efficient in turning
all neutron stars into  solar--mass black holes~--- after all,
thousands of  neutron stars are observed in our Galaxy. One way to
suppress their transmutation is to make the  
DM--neutron coupling moderately small, so that only the densest DM
environments would allow the stars to  accumulate the critical 
mass for collapse. Using the Milky Way parameters,
we require ${M_{\rm tot}^{\mathrm{MW}} \lesssim m N_{cr}}$ or
\begin{equation}
  \label{eq:13}
  y \lesssim 2\times 10^{-5} \; \left( m \over 100 \,
  \mbox{GeV}\right)^{1/2} \, \beta^{-1/2}\,,
\end{equation}
where Eqs.~\eqref{eq:total_capturedMW}, \eqref{eq:38},
and~\eqref{eq:69} were used. Note that Eq.~\eqref{eq:13} is not far
from the conditions~\eqref{eq:6}, \eqref{eq:12}; this is due to the fact that the
  DM parameters vary by only a few orders of magnitude from galaxy to
galaxy.

It is worth noting that we  have disregarded several
points which are important for the 
scenario of this paper. First, it is crucial that the DM is
non--annihilating and satisfies constraints coming
from its generation in the early Universe and from the DM detection
experiments. It remains to be seen if all these conditions can be
fitted together with the requirements of our mechanism.

Second, we have not analyzed the astrophysical signatures of
the neutron stars converting into the solar--mass black
holes. In our scenario, the transmutation
is controlled by a single parameter~---  the number of DM
particles $N_{cr}$ required for collapse. Thus, regardless of the
  underlying DM model the neutron stars will be converted in the parts of the Universe where   the DM
  abundance and velocity ensure its efficient accumulation. As a
  consequence,  solar--mass black holes will be
  distributed in a specific way among the galaxies of   different types. For
  example, a discovery of  old neutron stars in the dwarf galaxies~--- the
  best--known environments for the DM accumulation~--- would exclude
  a sizable overall abundance of the transmuted neutron stars in the Universe.

Third, several parts of our mechanism deserve a 
detailed numerical study. We expect that the phase transition of
the DM
particles into the $\chi$ quanta proceeds in a spectacular first--order way,
starting as a self--similar bosenova collapse~\cite{Zakharov12,
  Levkov:2016rkk} and ending up with  formation of a dense $\chi$
condensate. This full two--stage proscess has never been simulated
before; in the main text we have just crudely estimated the time
of $\chi$ condensation to be small, see Eq.~(\ref{eq:68}). The other
unexplored subject is the growth of the $\chi$ Bose--Einstein
condensate in the end of the process. We assumed that  the growth
continues even when the  condensate becomes incredibly small in
size. On the one hand, this optimism is based  on
the fact   that unlike  in Refs.~\cite{Levkov:2018kau,
  Eggemeier:2019jsu, Chen:2020cef,   Chen:2021oot}, the DM
interactions in our model are short--range. On the other hand,
 even in the worst case the growth of the
$\chi$--condensate may proceed in a recurring indirect way: by growing 
the $\phi$ condensate via thermalization to the point when the
transition $\phi \phi \to \chi$ happens, and then growing the $\phi$
condensate, again. The details of this process also should be studied
numerically.

Fourth, although viability of our new mechanism does not
  depend on the details of dark matter capture and thermalization, its
  predictions in a particular model rely on a quantitative description
  of these phenomena. Thus, precise identification of 
  a phenomenologically acceptable parameter region for the dark matter model should
  include studies of astrophysical uncertainties, equation of state
  for the nuclear matter, and DM interactions with it. Besides,
  renormalization group equations should be solved for the dark matter
  constants $\lambda'$ and $\lambda$ which are scale--dependent and
  run on par with  the Standard Model couplings.

Fifth and finally, it might be interesting to explore dark matter
  collapse in the centers of white dwarfs within our   model. Despite 
  being more dilute than the neutron stars, the latter objects are also
  better understood, which makes them a perspective testing
  ground for dark matter studies.

In a nutshell, our study demonstrates that almost any model of bosonic
DM can be modified at strong fields in such a way that the solar--mass
black holes can appear by transmuting the neutron stars. This calls
for a proper identification of low--mass compact objects in
astrophysical observations, cf.~\cite{LIGOScientific:2021psn}
and~\cite{LIGOScientific:2021iyk}.

\acknowledgments

The authors are indebted to Yoann G\'enolini and Thomas Hambye for
participation at the early stages of this project, and to Sergei Demidov and
Sebastien Clesse for discussions. Studies of neutron star environments
were funded by the Ministry of Science and Higher Education of the
Russian Federation under the state contract 075--15--2020--778
(project “Science”). The work of P.T. is supported in part by the IISN
grant 4.4503.15. RG is supported by MIUR grant PRIN 2017FMJFMW and
acknowledges the Galileo Galilei Institute for hospitality during this
work. DL thanks Universit\'{e} Libre de Bruxelles for hospitality. 

\appendix
\section{Q--ball and Bose star parameters}
\label{sec:bose-star-conf}

Here we derive parametric dependence of the soliton mass $M$ and
radius $R$ on its global charge $N$ in the models of
Sec.~\ref{sec:grow-bose-einst}. To this end we adopt a simple version
of the variational ansatz~\cite{Ho:1999hs, Chavanis:2011zi,
  Eby:2015hsq} ignoring order--one numerical coefficients wherever
possible. We will separately study the solitonic 
Q--ball~\cite{Nugaev:2019vru} stabilized by the attractive
self--interactions and the gravitationally bound Bose
star~\cite{Visinelli:2021uve} in the case of self--repulsion. 

Start with the Q--ball in the model (\ref{eq:54}) for the complex scalar
field $\phi(x)$. We will assume that its scalar potential $V$ becomes
flat, i.e.\ attractive, at strong fields ${|\phi|\gtrsim \Lambda_{\rm
    att}}$, where it can be roughly approximated as a 
power--law~(\ref{eq:57}) with $\alpha < 2$.  Suppose  the 
stationary Q--ball in this model has the form of a single bell--shaped 
lump with typical field strength $\varphi_0$ and size~$R$: 
\begin{equation}
  \label{eq:14}
  \phi = \varphi_0\,  f(\boldsymbol{x}/R) \, \mathrm{e}^{-i\omega t}\;,
\end{equation}
where $f(\boldsymbol{y})$ is a dimensionless order--one function with
the support at~${|\boldsymbol{y}| =  |\boldsymbol{x}|/R\lesssim 
  1}$;  the time dependence follows from Eq.~(\ref{eq:53}). 

Since the Q--ball is mostly bound by the self--attraction, we can
compute its parameters in flat spacetime ignoring
gravity. Substituting Eq.~(\ref{eq:14}) into the energy and
charge of this object, we get,
\begin{gather}
  M \sim \omega^2 \varphi_0^2 R^3 + \varphi_0^2 R
+ m^2 \varphi_0^\alpha \Lambda_{\rm att} ^{2-\alpha}R^3\;,
\label{eq:Eatt}\\[1px]
N \sim \omega \varphi_0^2 R^3\;,
\label{eq:Natt}
\end{gather}
where we omitted $\varphi_0$-- and $R$--independent numerical
coefficients of order~1 in front of every term~--- these come 
from the integrals over $\boldsymbol{y}$ of powers of
$f(\boldsymbol{y})$ and its derivatives. 

We now minimize $M$ with respect to $\varphi_0$ and $R$ at
a fixed~$N$. Expressing $\omega$ from Eq.~(\ref{eq:Natt}), we substitute
it into Eq.~(\ref{eq:Eatt}) and differentiate the result with respect
to the unknowns. We get,
\begin{align}
\notag
\varphi_0 {\partial M\over \partial \varphi_0} & \sim 
-{2 N^2\over \varphi_0^2 R^3} + 2 \varphi_0^2 R + \alpha m^2 \varphi_0^\alpha
\Lambda_{\rm att} ^{2-\alpha}R^3 = 0\;,\\
\notag
R\, {\partial M\over \partial R} & \sim 
- {3N^2\over \varphi_0^2 R^3}  + \varphi_0^2 R + 
3 m^2 \varphi_0^\alpha \Lambda_{\rm att} ^{2-\alpha}R^3 =0\;.
\end{align}
Notably, the solution of these equations does not exist at ${\alpha
  \geq 2}$ when self--attraction changes to self--repulsion. Away
from that  region, all terms in the equations are of the same order:
\begin{equation}
\label{eq:phi}
\varphi_0^{2-\alpha} \sim (mR)^2 \Lambda_{\mathrm{att}}^{2-\alpha}
\quad \mbox{and} \quad
N \sim \varphi_0^2R^2\,. 
\end{equation}
Using Eq.~(\ref{eq:Natt}), we find that ${\omega\sim R^{-1}}$ and
reproduce Eqs.~(\ref{eq:58}), (\ref{eq:59}) from the main text.

It is worth noting that $\omega \sim R^{-1}\ll m$ at sufficiently
large $N$, see Eq.~(\ref{eq:59}). Thus, the particles inside the large
Q--ball are very light due to mass deficit. As a consequence, the
overall soliton is lighter and less amenable to  gravitational
collapse than a collection of free particles with the same charge.

Now, consider a Bose star in the model with the scalar potential
(\ref{eq:61}) and $\alpha>2$. Recall that the flat--space solution 
does not  exist in this case~--- hence, we add  gravitational
attraction. In the stationary non--relativistic limit, this amounts
to introducing the metric $g_{00} = 1+2U$ and   $g_{ij} =
-(1-2U)\,\delta_{ij}$, where $U < 0$ is a time--independent Newtonian
potential, ${|U|\ll 1}$. With gravity included, the  energy and charge
of the nonrelativistic soliton read,
\begin{align}
\notag
&M =\int d^3\boldsymbol{x}\, \Bigl\{
(1-4U)|\partial_0\phi|^2 + |\partial_i\phi|^2  + (\partial_i
U)^2/(8\pi G) \\
&\qquad\qquad \;\;  + (1-2U)\, m^2 |\phi|^2 + m^2 |\phi|^\alpha
\Lambda_{\mathrm{rep}}^{2-\alpha} 
\Bigr\}\,,
\label{eq:Erep}
\\
&N = i\int d^3\boldsymbol{x}\, (1-4U) (\phi^* \partial_0\phi - \phi
\partial_0\phi^*)\,,
\label{eq:Nrep}
\end{align}
where the last term in the first line represents the energy density of the
gravitational field and we ignored the gravitational
effect of the self--interaction
energy~${\delta V \propto |\phi|^\alpha}$.

Following the same strategy as before, we assume that the solution has
the form~\eqref{eq:14} characterized by the field strength 
$\varphi_0$, size $R$, and the gravitational potential in the center
$U_0$. Omitting all order--one coefficients, again, we find, 
\begin{align}
\notag
& M \sim (1-4U_0)\, \omega^2 \varphi_0^2 R^3 + \varphi_0^2 R
+ U_0^2R / G\\
\label{eq:EE}
 & \quad  + (1-2U_0)\,  m^2 \varphi_0^2 R^3 +  m^2 \varphi_0^\alpha 
\Lambda_{\mathrm{rep}}^{2-\alpha} R^3\,,\\ 
\label{eq:NN}
& N \sim (1-4U_0)\, \omega \varphi_0^2 R^3\,,
\end{align}
cf.\ Eqs.~\eqref{eq:Eatt}, \eqref{eq:Natt}. Below we also
disregard the gradient energy of the soliton $\varphi_0^2 R$,
checking a posteriori that it is small.

Next, we express $\omega$
from Eq.~(\ref{eq:NN}), substitute it into Eq.~(\ref{eq:EE}), and
minimize the resulting energy with respect to $\varphi_0$, $R$, and
$U_0$. The extremality conditions are,
\begin{align}
\notag
& - (1+4U_0)\, {2N^2\over \varphi_0^2R^3} + 2\, (1-2U_0)\, m^2
\varphi_0^2 R^3 \\
\label{eq:eqphi}
&\mbox{\hspace{3.7cm}}+ \alpha m^2 \varphi_0^\alpha 
\Lambda_{\mathrm{rep}}^{2-\alpha}R^3 = 0\,,\\
\notag
& -(1+4U_0)\, {3 N^2\over \varphi_0^2R^3} + 3 \,(1-2U_0)\, m^2
\varphi_0^2 R^3 \\
\label{eq:eqR}
&\mbox{\hspace{2cm}} + U_0^2R / G + 3 m^2 \varphi_0^\alpha
\Lambda_{\mathrm{rep}}^{2-\alpha}R^3 = 0\,,\\
\label{eq:15}
& \frac{2N^2}{\varphi_0^2 R^3} + \frac{U_0R}G - m^2 \varphi_0^2 R^3 = 0\;,
\end{align}
where we recalled that $|U_0| \ll 1$. It is peculiar that the above system
is almost degenerate: Eqs.~\eqref{eq:eqphi} and \eqref{eq:eqR} differ
only by the last terms suppressed as $U_0^2$ or
$\Lambda_{\mathrm{rep}}^{2 - \alpha}$. This is a consequence of the
fact that the dominant part of the potential is quadratic. To the
leading order, both equations give the standard expression for the
number of nonrelativistic particles, 
\begin{equation}
\label{eq:leading}
N \sim m \varphi_0^2 R^3\,. 
\end{equation}
Now, Eq.~(\ref{eq:15}) takes a familiar form ${U_0 \sim -GmN/R}$.  The
last relation is found from the linear combination of
Eqs.~\eqref{eq:eqphi} and~\eqref{eq:eqR} such that the large terms
  cancel. We get,
\begin{equation}
\label{eq:virial}
2U_0^2/G \sim 3\, (\alpha-2)\, (mR)^2
\varphi_0^\alpha\Lambda_{\mathrm{rep}}^{2-\alpha} \,. 
\end{equation}
Since the left--hand side is positive, the solution exists precisely
in the case~$\alpha>2$. One finally obtains,
\begin{equation}
\label{eq:phi-rep}
\varphi_0 \sim \Lambda_{\mathrm{rep}} \; \xi^{\,\mbox{\footnotesize $2 \over 3\alpha-8$}}\;,
\qquad  R \sim  \frac{M_{pl}}{m\Lambda_{\mathrm{rep}}}\; \xi
^{\,\mbox{\footnotesize $\alpha-4 \over 3\alpha - 8$}}\,,
\end{equation}
where $\xi \equiv N m^2\Lambda_{\mathrm{rep}}/M_{pl}^3$.

The interpretation of the above solution is essentially
different in the cases $2 < \alpha < 8/3$ and ${\alpha > 8/3}$.  If 
$\alpha$ exceeds $8/3$, the Bose star is stable at $\xi < 1$. Its field
steadily grows with the multiplicity reaching  ${\varphi_0 \sim
\Lambda_{\mathrm{rep}}}$ at $\xi \sim 1$. At this point, the soliton's
gravitational potential $U_0 \sim  - \xi^{(2\alpha -
  4)/(3\alpha - 8)}$ becomes of order one and a black hole forms,
see Eq.~(\ref{eq:virial}). Notably, at the brink of 
collapse $mR \sim M_{pl}/\Lambda_{\mathrm{rep}} \gg 1$ --- thus, the
soliton remains nonrelativistic, indeed. Also, the gradient term
$\varphi_0^2 R$ in Eq.~\eqref{eq:EE} is much smaller than the binding
energy $U_0^2 R/G$, as we assumed.

In the opposite case $2< \alpha < 8/3$ the scalar self--repulsion
triumphs over gravity at large distances. Indeed, expressing
$\varphi_0$ from Eq.~(\ref{eq:leading}), one learns that at a fixed $N$ 
the self--repulsion energy of the nonrelativistic condensate grows faster with
its size $R$ than the gravitational binding energy:
$(\varphi_0^\alpha R^3)/(U_0^2 R)\propto R^{(8 -   3\alpha)/2}$, see
Eq.~(\ref{eq:EE}). This means that clumps of  Bose--Einstein
condensate spread over the entire volume offered by the
external conditions. The solutions (\ref{eq:phi-rep}) in this case
break the Vakhitov--Kolokolov criterion~\cite{vk,   Zakharov12}
$d\omega / dN \sim m \, dU_0 / dN < 0$ necessary for stability. Thus,
they represent the maxima of the  potential energy and determine the
minimum size $R$ to which the condensate should be
squeezed for gravity to dominate. We  do not consider any 
external forces in this paper~--- hence, the black hole does not form
in the case $2 < \alpha < 8/3$ at all.

\bibliography{NS}{} 

\end{document}